\documentstyle[12pt,fleqn]{article}

\def\thebibliography#1{\section*{References}\list
  {[\arabic{enumi}]}{\settowidth\labelwidth{#1}\leftmargin\labelwidth
    \advance\leftmargin\labelsep
    \usecounter{enumi}}
    \def\newblock{\hskip .11em plus .33em minus .07em}
    \sloppy\clubpenalty4000\widowpenalty4000
    \sfcode`\.=1000\relax}

\def\op#1{\mathop{\fam0 #1}\limits}

\newcommand{\beq}{\begin{equation}}
\newcommand{\eeq}{\end{equation}}
\newcommand{\ben}{\begin{eqnarray}}
\newcommand{\een}{\end{eqnarray}}
\newcommand{\be}{\begin{eqnarray*}}
\newcommand{\ee}{\end{eqnarray*}}
\newcommand{\bea}{\begin{eqalph}}
\newcommand{\eea}{\end{eqalph}}

\newcommand{\cA}{{\cal A}}

\newcommand{\cD}{{\cal D}}

\newcommand{\cH}{{\cal H}}

\newcommand{\bL}{{\bf L}}
\newcommand{\bR}{{\bf R}}
\newcommand{\bC}{{\bf C}}
\newcommand{\bZ}{{\bf Z}}

\newcommand{\bT}{{\bf T}}

\newcommand{\la}{\lambda}

\newcommand{\f}{\phi}
\newcommand{\om}{\omega}
\newcommand{\Om}{\Omega}
\newcommand{\m}{\mu}

\newcommand{\g}{\gamma}
\newcommand{\G}{\Gamma}

\newcommand{\vt}{\vartheta}
\newcommand{\up}{\upsilon}
\newcommand{\lng}{\langle}
\newcommand{\rng}{\rangle}

\newcommand{\w}{\wedge}
\newcommand{\wt}{\widetilde}
\newcommand{\wh}{\widehat}
\newcommand{\ol}{\overline}
\newcommand{\dr}{\partial}

\newcommand{\ot}{\otimes}

\newcounter{eqalph}
\newcounter{equationa}
\newcounter{theorem}
\newcounter{remark}
\newcounter{proposition}
\newcounter{lemma}
\newcounter{corollary}
\newcounter{definition}

\newenvironment{eqalph}{\stepcounter{equation}
\setcounter{equationa}{\value{equation}}
\setcounter{equation}{0}

\begin{eqnarray}}{\end{eqnarray}\setcounter{equation}{\value{equationa}}}

\def\theremark{\arabic{remark}}

\def\thedefinition{\arabic{definition}}

\newcommand{\mar}[1]{}

\begin{document}
\hbox{}

{\parindent=0pt

{\large\bf Covariant geometric quantization of non-relativistic
 Hamiltonian mechanics}
\bigskip

{\sc Giovanni Giachetta$^\dagger$\footnote{E-mail
address: giachetta@campus.unicam.it}, Luigi
Mangiarotti$^\dagger$\footnote{E-mail address: mangiaro@camserv.unicam.it} and
Gennadi  Sardanashvily$^\ddagger$\footnote{E-mail address:
sard@grav.phys.msu.su}}
\bigskip

\begin{small}
$\dagger$ Department of Mathematics and Physics, University of Camerino, 62032
Camerino (MC), Italy \\
$\ddagger$ Department of Theoretical Physics, Physics Faculty, Moscow State
University, 117234 Moscow, Russia
\bigskip

{\bf Abstract.}
We provide geometric quantization of the vertical cotangent bundle $V^*Q$
equipped with the canonical Poisson structure. This is a momentum phase space
of non-relativistic mechanics with the configuration bundle $Q\to\bR$. The
goal is the Schr\"odinger representation of $V^*Q$. We show that this
quantization is equivalent to the fibrewise quantization of symplectic fibres
of $V^*Q\to\bR$ that makes the quantum algebra of non-relativistic mechanics 
an instantwise algebra. Quantization of the classical evolution equation
defines a connection on this instantwise algebra, which provides quantum
evolution in non-relativistic mechanics as a parallel transport along time.  
\end{small}
}


\section{Introduction}

We study covariant geometric quantization of non-relativistic Hamiltonian
mechanics subject to time-dependent transformations. 

Its configuration space is a fibre bundle $Q\to
\bR$ equipped with bundle coordinates 
$(t,q^k)$, $k=1,\ldots,m$, where $t$ is the Cartesian coordinate on the
time axis $\bR$ with affine transition functions $t'=t+$const. 
Different trivializations
$Q\cong
\bR\times M$ of $Q$ correspond to different non-relativistic reference frames.
In contrary to all the existent quantizations of non-relativistic mechanics
(e.g., \cite{sni,wood}), we do not fix a trivialization of $Q$.
 
The momentum phase space of non-relativistic mechanics is the vertical
cotangent bundle
$V^*Q$ of $Q\to\bR$, 
endowed with the holonomic coordinates $(t,q^k,p_k)$.
It is provided with the canonical Poisson structure 
\mar{m72}\beq
\{f,f'\}_V = \dr^kf\dr_kf'-\dr_kf\dr^kf', \qquad f,f'\in C^\infty(V^*Q),
\label{m72}
\eeq
whose symplectic foliation coincides with the fibration $V^*Q\to\bR$
\cite{book98,sard98}. 
Given a trivialization
\mar{gm156}\beq
V^*Q\cong\bR\times T^*M, \label{gm156}
\eeq
the Poisson manifold $(V^*Q,\{,\}_V)$ is isomorphic to the direct
product of the Poisson manifold $\bR$ with the zero Poisson structure and the
symplectic manifold $T^*M$. 
An important peculiarity of the Poisson structure (\ref{m72}) is
that the Poisson algebra $C^\infty(V^*Q)$ of smooth real functions on $V^*Q$
is a Lie algebra over the ring $C^\infty(\bR)$ of functions of time alone.

Our goal is the geometric quantization of the Poisson bundle $V^*Q\to\bR$,
but it is not sufficient for quantization of non-relativistic mechanics.

The
problem is that non-relativistic mechanics can not be described 
as a Poisson Hamiltonian system on the momentum phase space $V^*Q$. Indeed, a
non-relativistic Hamiltonian $\cH$ is not an element of the Poisson algebra
$C^\infty(V^*Q)$. Its definition involves the cotangent bundle $T^*Q$ of $Q$.
Coordinated by $(q^0=t,q^k,p_0=p,p_k)$, the cotangent bundle
$T^*Q$ plays the role of the homogeneous momentum phase space of
non-relativistic mechanics. It
is equipped with the canonical Liouville form $\Xi=p_\la dq^\la$, the
symplectic form
$\Om=d\Xi$, and the corresponding Poisson bracket  
\be
\{f,g\}_T =\dr^\la f\dr_\la f -\dr_\la
f\dr^\la f', \qquad f,f'\in C^\infty(T^*Q).
\ee
Due to the one-dimensional canonical fibration
\mar{z11}\beq
\zeta:T^*Q\to V^*Q, \label{z11}
\eeq
the cotangent bundle $T^*Q$ provides the symplectic realization of the Poisson
manifold
$V^*Q$, i.e., 
\be
\zeta^*\{f,f'\}_V=\{\zeta^*f,\zeta^*f'\}_T 
\ee
for all $f,f'\in C^\infty(V^*Q)$.  
A Hamiltonian on $V^*Q$ is defined as 
 a global section 
\mar{qq4}\beq
h:V^*Q\to T^*Q, \qquad p\circ
h=-\cH(t,q^j,p_j), \label{qq4}
\eeq
of the one-dimensional affine bundle (\ref{z11}) \cite{book98,sard98}. 
As a consequence (see Section 6), the evolution equation of non-relativistic
mechanics is expressed into the Poisson bracket $\{,\}_T$ on $T^*Q$. It reads
\mar{mm17}\beq
\vt_{H^*}(\zeta^*f) =\{\cH^*,\zeta^*f\}_T, \label{mm17}
\eeq
where $\vt_{H^*}$ is the Hamiltonian vector field of the function
\mar{mm16}\beq
\cH^*=\dr_t\rfloor(\Xi-\zeta^* h^*\Xi))=p+\cH \label{mm16}
\eeq
on $T^*Q$.

Therefore, we need the compatible geometric quantizations 
both of the cotangent bundle  $T^*Q$ and
the vertical cotangent bundle $V^*Q$ such that 
the  monomorphism
\mar{qq60}\beq
\zeta^*: (C^\infty(V^*Q),\{,\}_V) \to (C^\infty(T^*Q),\{,\}_T) \label{qq60}
\eeq
of the Poisson algebra on $V^*Q$ to that on $T^*Q$ is prolonged to a
monomorphism of quantum algebras of $V^*Q$ and $T^*Q$.

Recall that  the
geometric quantization procedure falls into  three
steps: prequantization, polarization and metaplectic correction (e.g.,
\cite{eche98,sni,wood}).  Given a symplectic  manifold $(Z,\Om)$ and the
corresponding Poisson bracket $\{,\}$, prequantization associates to each
element $f$ of the Poisson algebra
$C^\infty(Z)$ on $Z$ 
a first order differential operator
$\wh f$ in the space of sections of a complex line bundle $C$ over $Z$
such that the Dirac condition
\mar{qm514}\beq
[\wh f,\wh f']=-i\wh{\{f,f'\}} \label{qm514}
\eeq
holds. 
Polarization of a symplectic manifold $(Z,\Om)$
is defined as a maximal involutive distribution $\bT\subset TZ$ such that 
Orth$_\Om \bT=\bT$, i.e., 
\mar{qq26}\beq
\Om(\vt,\up)=0, \qquad \forall \vt,\up\in\bT_z, \qquad z\in Z. \label{qq26}
\eeq
Given the Lie algebra $\bT(Z)$ of global sections of 
${\bf T}\to Z$, let $\cA_T\subset C^\infty(Z)$ denote the subalgebra of 
 functions $f$ whose Hamiltonian vector fields $\vt_f$ fulfill the condition
\mar{qq105}\beq
[\vt_f,\bT(Z)]\subset \bT(Z). \label{qq105}
\eeq
Elements of this subalgebra are only quantized. Metaplectic
correction provides the pre-Hilbert space $E_T$ where the quantum algebra
$\cA_T$ acts by symmetric operators. This is a certain subspace of sections of
the tensor product $C\ot\cD_{1/2}$ of the prequantization line bundle $C\to
Z$ and a bundle $\cD_{1/2}\to Z$ of half-densities on $Z$.  
The
geometric quantization  procedure has been extended to Poisson manifolds
\cite{vais91,vais} and to Jacobi manifolds
\cite{leon}.

We show that standard prequantization of the cotangent bundle
$T^*Q$ (e.g., \cite{eche98,sni,wood}) provides the 
compatible prequantization of the Poisson manifold $V^*Q$ such that
the monomorphism
$\zeta^*$ (\ref{qq60}) is prolonged to a monomorphism of prequantum algebras.

In contrast with the prequantization procedure, polarization of $T^*Q$ need
not imply a compatible polarization of $V^*Q$, unless it includes the vertical
cotangent bundle $V_\zeta T^*Q$ of the fibre bundle $\zeta$ (\ref{z11}), i.e.,
spans over vectors
$\dr^0$. 
The canonical real polarization of $T^*Q$, satisfying 
the condition 
\mar{qq14}\beq
V_\zeta T^*Q\subset \bT, \label{qq14}
\eeq
is the vertical polarization. It
coincides with the vertical tangent bundle $VT^*Q$ of $T^*Q$, i.e., spans over
all the vectors
$\dr^\la$. We show that this polarization and the corresponding metaplectic
correction of $T^*Q$ induces the compatible quantization of the Poisson
manifold
$V^*Q$  such that the monomorphism of Poisson algebras $\zeta^*$ (\ref{qq60})
is prolonged to a monomorphism of quantum algebras of $V^*Q$ and $T^*Q$. The
quantum algebra
$\cA_V$ of $V^*Q$ consists of functions on $V^*Q$ which are at most
affine in momenta $p_k$ and have the Schr\"odinger
representation in the space of half-densities on $Q$. It is essential
that, since these operators does not contain the derivative with respect to
time, the quantum algebra
$\cA_V$ is a $C^\infty(\bR)$-algebra.

We prove that the Schr\"odinger quantization of $V^*Q$ yields geometric
quantization of symplectic fibres of
the Poisson bundle $V^*Q\to\bR$ such that any quantum operator $\wh f$ on
$V^*Q$, restricted to the fibre
$V^*_tQ$, $t\in \bR$, coincides with the quantum operator on
the symplectic manifold $V^*_tQ$ of the function $f|_{V^*Q}$. 
Thus, the quantum algebra $\cA_V$ of the Poisson bundle $V^*Q\to\bR$
can be seen as the
instantwise $C^\infty(\bR)$-algebra of its symplectic fibres. This agrees
with the instantwise quantization of 
symplectic fibres $\{t\}\times T^*M$ of the
direct product (\ref{gm156}) in \cite{sni}.

A fault of the Schr\"odinger representation on half-densities is that the
Hamiltonian function $\cH^*$ (\ref{mm16}) does not belong to the quantum
algebra $\cA_T$ of $T^*Q$ in general. A rather sophisticated solution of this
problem for quadratic Hamiltonians has been suggested in 
\cite{sni}. However, the Laplace operator constructed in
\cite{sni} does not fulfill the Dirac condition (\ref{qm514}).
If a Hamiltonian $\cH$ is a polynomial of momenta $p_k$, one can
represent it as an element of the universal enveloping algebra of the Lie
algebra $\cA_T$, but this representation is not necessarily globally defined.

In order to include a Hamiltonian function $\cH^*$ (\ref{mm16}) to the quantum
algebra,  one can choose the
Hamiltonian polarization of $T^*Q$ which contains the Hamiltonian vector field
$\vt_{\cH^*}$ of $\cH^*$. However, it
does not satisfy the condition (\ref{qq14}) and does not define any
polarization of the Poisson manifold $V^*Q$. This polarization is the necessary
ingredient in a different variant of geometric quantization of
$V^*Q$ which is seen as a presymplectic manifold $(V^*Q,h^*\Om)$. Given a
trivialization (\ref{gm156}), this quantization has been studied in
\cite{wood}. In Section 5, its frame-covariant form is discussed.

Finally, since the quantum algebra $\cA_V$ of the Poisson manifold $V^*Q$ is a
$C^\infty(\bR)$-algebra and since $\wh p=-i\dr/\dr t$, quantization of the
classical evolution equation (\ref{mm17}) defines the connection
\mar{qq120}\beq
\nabla\wh f= i[\wh \cH^*,\wh f] \label{qq120}
\eeq
on the enveloping quantum algebra $\ol\cA_V$ and describes quantum evolution 
in non-relativistic mechanics as a parallel transport along time.

\section{Prequantization}

Basing on the standard prequantization of the cotangent bundle $T^*Q$,
we here construct the compatible prequantizations of the Poisson
bundle $V^*Q\to\bR$ and its symplectic leaves. 

Recall the prequantization of $T^*Q$ (e.g., \cite{eche98,sni,wood}). 
Since its symplectic form $\Om$ is exact and belongs to the zero de
Rham cohomology class,  
the prequantization bundle is the trivial complex line bundle 
\mar{qm501}\beq
C=T^*Q\times\bC\to T^*Q,  \label{qm501}
\eeq
whose Chern class $c_1$ is zero. Coordinated by $(q^\la,p_\la,c)$,
it is provided with the
admissible linear connection
\mar{qm502}\beq
A=dp_\la\ot\dr^\la +dq^\la\ot(\dr_\la+ ip_\la c\dr_c) \label{qm502}
\eeq
with the strength form $F= - i\Om$ and the
Chern form 
\be
c_1=\frac{i}{2\pi}F=\frac1{2\pi}\Om.
\ee
The $A$-invariant Hermitian fibre metric on $C$ is
$g(c,c)=c\ol c$.
The covariant derivative of sections $s$ of the prequantization bundle $C$
(\ref{qm501}) relative to the connection $A$ (\ref{qm502}) along the vector
field
$u$ on $T^*Q$ takes the form
\mar{qq12}\beq
\nabla_u(s)= (u^\la\dr_\la - iu^\la p_\la)s. \label{qq12}
\eeq
Given a function $f\in C^\infty(T^*Q)$, 
the covariant derivative (\ref{qq12}) along 
the Hamiltonian vector field 
\be
\vt_f=\dr^\la f\dr_\la -\dr_\la f\dr^\la, \qquad \vt_f\rfloor\Om=-df
\ee
of $f$ reads
\be
\nabla_{\vt_f}= \dr^\la f(\dr_\la - ip_\la) - \dr_\la f\dr^\la.
\ee
Then, in order to satisfy the Dirac condition (\ref{qm514}), one
assigns to each element
$f$ of the Poisson algebra 
$C^\infty(T^*Q)$ the first order differential operator 
\mar{qm504}\beq
\wh f(s) =-i(\nabla_{\vt_f} + if)s=[-i\vt_f+ (f -p_\la\dr^\la f)]s
\label{qm504}
\eeq
on sections
$s\in C(T^*Q)$ of the prequantization bundle $C$ (\ref{qm501}).
For instance, the prequantum operators (\ref{qm504}) for local functions 
$f=p_\la$, $f=q^k$, a global function $f=t$, and the constant function
$f= 1$  read
\be
\wh p_\la= -i\dr_\la, \qquad \wh q^\la =i \dr^\la +  q^\la,
\qquad \wh 1= 1. 
\ee
For elements $f$ of the Poisson subalgebra $C^\infty(V^*Q)\subset
C^\infty(T^*Q)$, the Kostant--Souriau
formula (\ref{qm504}) takes the form
\mar{qq40}\beq
\wh f(s) =[-i(\dr^kf\dr_k-\dr_\la f\dr^\la)+ (f -p_k\dr^k f)]s.
\label{qq40}
\eeq

Turn now to
prequantization of the Poisson manifold $(V^*Q,\{,\}_V)$.  
The Poisson bivector $w$ of the Poisson structure on
$V^*Q$ reads
\mar{qq41}\beq
w=\dr^k\w\dr_k=-[w,u]_{\rm SN}, \label{qq41}
\eeq
where $[,]_{\rm SN}$ is the Schouten--Nijenhuis bracket and $u=p_k\dr^k$ is
the Liouville vector field on the vertical cotangent bundle $V^*Q\to Q$.
The relation (\ref{qq41}) shows that the Poisson bivector $w$ is
exact and, consequently, has the zero Lichnerowicz--Poisson cohomology
class
\cite{book98,vais}. Therefore, let us consider the trivial complex line bundle 
\mar{qq43}\beq
C_V=V^*Q\times \bC\to V^*Q \label{qq43}
\eeq
such that the zero Lichnerowicz--Poisson cohomology class of $w$ is the image
of the zero Chern class $c_1$ of $C_V$ under the cohomology homomorphisms 
\be
H^*(V^*Q,\bZ) \to H^*_{\rm deRh}(V^*Q) \to H^*_{\rm LP}(V^*Q).
\ee

Since the line bundles $C$ (\ref{qm501}) and $C_V$ (\ref{qq43}) are trivial, 
$C$
can be seen as the pull-back
$\zeta^*C_V$ of $C_V$, while $C_V$ is isomorphic to the
pull-back
$h^*C$ of $C$ with respect to a section 
$h$ (\ref{qq4}) of the affine bundle (\ref{z11}).   
Since $C_V=h^*C$ and since the covariant derivative of the connection $A$
(\ref{qm502}) along the fibres of $\zeta$ (\ref{z11}) is trivial, let us
consider the pull-back 
\mar{qq44}\beq
h^*A=dp_k\ot\dr^k +dq^k\ot(\dr_k+ ip_k c\dr_c)+ 
dt\ot(\dr_t - i\cH c\dr_c) \label{qq44}
\eeq
of the connection $A$ (\ref{qm502}) onto $C_V\to V^*Q$
\cite{book00}.
This connection defines the
contravariant derivative 
\mar{qq45}\beq
\nabla_\f s_V = \nabla_{w^\sharp\f}s_V \label{qq45}
\eeq
of sections $s_V$ of $C_V\to V^*Q$ along one-forms $\f$ on $V^*Q$,
which corresponds to a contravariant connection $A_V$ on the line bundle
$C_V\to V^*Q$ \cite{vais}. It is readily observed that this
contravariant connection does not depend on the choice of a section $h$. 
By virtue of the relation (\ref{qq45}), the curvature bivector of $A_V$ equals
to
$-iw$  \cite{vais97}, i.e., $A_V$ is an admissible connection
for the canonical Poisson structure on $V^*Q$. Then the 
 Kostant--Souriau formula 
\mar{qq46}\beq
\wh f_V(s_V) =(-i\nabla_{\vt_{Vf}}+f)s_V=[-i(\dr^kf\dr_k-\dr_kf\dr^k)+ (f
-p_k\dr^k f)]s_V
\label{qq46}
\eeq
defines prequantization of the Poisson manifold $V^*Q$. 

In particular, the prequantum operators of
functions   
$f\in C^\infty(\bR)$ of time alone reduces simply to multiplication  $\wh
f_Vs_V=fs_V$ by these functions. Consequently, the prequantum algebra
$\wh C^\infty(V^*Q)$ inherits the structure of a $C^\infty(\bR)$-algebra.

It is immediately observed that the prequantum operator $\wh f_V$
(\ref{qq46}) coincides with the prequantum operator $\wh{\zeta^*f}$
(\ref{qq40}) restricted to the pull-back sections $s=\zeta^*s_V$ of the line
bundle $C$. Thus, prequantization of the Poisson algebra $C^\infty(V^*Q)$ on
the Poisson manifold $(V^*Q,\{,\})$ is equivalent to its prequantization as a
subalgebra of the Poisson algebra $C^\infty(T^*Q)$ on the symplectic 
manifold $T^*Q$.

The above prequantization of the Poisson manifold $V^*Q$ yields
prequantization of its symplectic leaves as follows. 

Since $w^\sharp\f=\f^k\dr_k-\f_k\dr^k$
is a vertical vector field
on $V^*Q\to\bR$ for any one-form $\f$ on $V^*Q$, the contravariant derivative
(\ref{qq45}) defines a connection along each fibre $V_t^*Q$, $t\in\bR$, of the
Poisson bundle
$V^*Q\to\bR$. This is the pull-back  
\be
A_t=i^*_t h^*A= dp_k\ot\dr^k +dq^k\ot(\dr_\la+ ip_k c\dr_c)
\ee
of the connection
$h^*A$ (\ref{qq44}) on the
pull-back bundle $i_t^*C_V\to V_t^*Q$ with respect to the imbedding
$i_t:V^*_tQ\to V^*Q$. It is readily observed that this connection is
admissible for the symplectic structure 
\be
\Om_t=dp_k\w dq^k
\ee
 on $V^*_tQ$, and
provides prequantization of the symplectic manifold $(V_t^*Q,\Om_t)$. The
corresponding prequantization formula is given by the expression
(\ref{qq46}) where functions $f$ and sections $s_V$ are restricted to
$V^*_tQ$. 
Thus, the prequantization (\ref{qq46}) of the Poisson manifold
$V^*Q$ is a leafwise 
prequantization \cite{vais97}. 

\section{Polarization}

Given compatible prequantizations of the cotangent bundle $T^*Q$, the Poisson
bundle $V^*Q\to \bR$ and its simplectic fibres, let us now construct their
compatible polarizations.  

Recall that, given a polarization $\bT$ of a prequantum symplectic manifold
$(Z,\Om)$, the subalgebra $\cA_T\subset C^\infty(Z)$ of 
functions $f$ obeying the condition
(\ref{qq105}) is only quantized.
Moreover, after further metaplectic correction, one consider
a representation of this algebra in a quantum space $E_T$ such
that
\mar{qq15}\beq
\nabla_u e =0, \qquad \forall u\in \bT(Z), \qquad e\in E_T. \label{qq15}
\eeq

Recall that by a polarization of a Poisson manifold $(Z,\{,\})$ is meant a
sheaf
$\bT^*$ of germs of complex functions on $Z$ whose stalks $\bT^*_z$, $z\in Z$,
are Abelian algebras with respect to the Poisson bracket $\{,\}$ \cite{vais97}.
One can also require that the algebras $\bT^*_z$ are maximal,
but this condition need not hold under pull-back and push-forward operations. 
Let $\bT^*(Z)$ be the structure
algebra  of global sections of the sheaf $\bT^*$; it is
also called a Poisson polarization \cite{vais91,vais}.
A quantum algebra $\cA_T$ associated to the Poisson polarization $\bT^*$ is
defined as a subalgebra of the Poisson algebra
$C^\infty(Z)$ which consists of functions $f$ such that
\be
\{f,\bT^*(Z)\}\subset \bT^*(Z).
\ee
Polarization of a symplectic manifold yields its maximal Poisson
polarization, and {\it vice versa}.

There are different polarizations of the cotangent bundle $T^*Q$. We will
consider those polarizations of $T^*Q$ whose direct image as Poisson
polarizations  onto
$V^*Q$ with respect to the morphism $\zeta$ (\ref{z11}) are polarizations of
the Poisson manifold $V^*Q$. This takes place if the germs of
polarization 
$\bT^*$ of
the Poisson manifold $(T^*Q,\{,\}_T)$ are constant along the fibres of the
fibration $\zeta$ (\ref{z11}) \cite{vais97}, i.e., are germs of functions
independent of the momentum coordinate $p_0=p$. It means that the corresponding
polarization
$\bT$ of the symplectic manifold
$T^*Q$ is vertical with respect to the fibration $T^*Q\to\bR$, i.e., obeys the
condition (\ref{qq14}). A short calculation shows that, in this case,
the associated quantum algebra
$\cA_T$ consists of functions $f\in C^\infty(T^*Q)$ which are at most affine in
the momentum coordinate $p_0$. Moreover, given such
a polarization, the equality (\ref{qq15}) implies the equality
\be
\nabla_{u_0\dr^0}e =0, \qquad e\in E_T,
\ee 
for any vertical vector field $u_0\dr^0$ on the fibre bundle $T^*Q\to V^*Q$. 
Then the prequantization formulas (\ref{qq40}) and (\ref{qq46}) for the
Poisson algebra $C^\infty(V^*Q)$ coincide on quantum spaces, i.e., the
monomorphism $\zeta^*$ (\ref{qq60}) is prolonged to monomorphism of quantum
algebras of $V^*Q$ and $T^*Q$. 

The vertical polarization $VT^*Q$ of $T^*Q$ obeys the
condition (\ref{qq14}). It
is a strongly admissible polarization, and its integral manifolds are fibres of
the cotangent bundle
$T^*Q\to Q$.
One can verify easily that the associated quantum
algebra $\cA_T$ consists of functions on $T^*Q$ which are at most affine in
momenta
$p_\la$. 
The quantum space $E_T$ associated to the vertical polarization obeys the
condition 
\mar{qq80}\beq
\nabla_{u^\la\dr_\la} e=0, \qquad \forall e\in E_T. \label{qq80}
\eeq
Therefore, the operators of the quantum algebra $\cA_T$ on this quantum space
read
\mar{qq20}\beq
f=a^\la(q^\mu)p_\la + b(q^\mu), \qquad \wh f= -i\nabla_{a^\la\dr_\la} +b.
\label{qq20}
\eeq
This is the Schr\"odinger representation of $T^*Q$.

The vertical polarization of $T^*Q$ defines the maximal polarization
$\bT^*$ of the Poisson manifold $V^*Q$ which consists of germs of functions
constant on the fibres of $V^*Q\to Q$. The
associated quantum space $E_V$ obeys the condition 
\mar{qq70}\beq
\nabla_{u^k\dr_k} e=0, \qquad \forall e\in E_V. \label{qq70}
\eeq
The quantum algebra $\cA_V$ corresponding to this polarization of
$V^*Q$ consists of functions on $V^*Q$ which are at most affine in momenta
$p_k$. Their quantum operators read
\mar{qq20'}\beq
f=a^\la(q^\mu)p_k + b(q^\mu), \qquad \wh f= -i\nabla_{a^k\dr_k} +b.
\label{qq20'}
\eeq
This is the Schr\"odinger representation of $V^*Q$.

In turn, each symplectic
fibre $V^*_tQ$, $t\in \bR$, of the Poisson bundle $V^*Q\to \bR$ is
provided with the pull-back polarization $\bT_t^*=i^*_t\bT^*$ 
with respect to the Poisson morphism $i_t:V^*_tQ\to V^*Q$. The corresponding
distribution $\bT_t$ 
coincides with the vertical tangent bundle of the fibre bundle $V^*_tQ\to
Q_t$. The associated quantum algebra $\cA_t$ consists of functions  on
$V^*Q_t$ which are at most affine in momenta
$p_k$, while the quantum space $E_t$ obeys
the condition similar to (\ref{qq70}). Therefore, the representation of the
quantum algebra $\cA_t$ takes the form (\ref{qq20'}). It follows that the
Schr\"odinger representation the Poisson bundle $V^*Q\to \bR$ 
is a fibrewise representation.  

\section{Metaplectic correction}

To complete the geometric quantization procedure of $V^*Q$,
let us consider the metaplectic correction of the
Schr\"odinger representations of $T^*Q$ and
$V^*Q$.

The representation of the quantum algebra $\cA_T$ (\ref{qq20}) can be
defined in the subspace of sections of the line bundle $C\to T^*Q$ which
fulfill the relation (\ref{qq80}). 
This representation reads
\mar{qq81}\beq
f=a^\la(q^\mu)p_\la + b(q^\mu), \qquad \wh f= -ia^\la\dr_\la +b.
\label{qq81}
\eeq
Therefore, it can be restricted to the sections $s$ of the pull-back
line bundle
$C_Q=\wh 0^*C\to Q$ where $\wh 0$ is the canonical zero section of
the cotangent bundle $T^*Q\to Q$.
However, this is not yet a representation
in a Hilbert space. 

Let $Q$ be an oriented manifold. Applying the general
metaplectic technique \cite{eche98,wood}, we come to the
vector bundle $\cD_{1/2}\to Q$ of complex half-densities on $Q$ with
the transition functions $\rho'=J^{-1/2}\rho$,
where $J$ is the Jacobian of the coordinate transition functions 
on $Q$. Since $C_Q\to Q$ is a trivial bundle, the tensor product
$C_Q\ot\cD_{1/2}$ is isomorphic to $\cD_{1/2}$. Therefore, the quantization
formula (\ref{qq81}) can be extended to sections of the half-density bundle
$\cD_{1/2}\to Q$ as 
\mar{qq82}\beq
f=a^\la(q^\mu)p_\la + b(q^\mu), \qquad
\wh f\rho=(-i\bL_{a^\la\dr_\la} +b)\rho= (-ia^\la\dr_\la 
-\frac{i}{2}\dr_\la (a^\la)+ b)
\rho,
\label{qq82}
\eeq
where $\bL$ denotes the Lie derivative.
The second term in the right-hand side of
this formula is a  metaplectic correction. It makes the operator $\wh f$
(\ref{qq82})  symmetric with respect to the Hermitian form
\be
\lng \rho_1|\rho_2\rng=\left(\frac{1}{2\pi}\right)^m
\op\int_Q \rho_1 \ol\rho_2
\ee
on the pre-Hilbert space $E_T$ of sections of
$\cD_{1/2}$ with compact support. The completion $\ol E_T$ of $E_T$ provides a
Hilbert space of the
Schr\"odinger representation of the quantum algebra $\cA_T$, where the
operators (\ref{qq82}) are essentially
self-adjoint, but not necessarily  bounded. Of course, functions of
compact support on the time axis $\bR$ have a limited physical application,
but we can always restrict our consideration to some bounded interval of
$\bR$. 

Since, in the case of the vertical polarization, there is a monomorphism of
the quantum algebra $\cA_V$ to the quantum algebra $\cA_T$, one can define
the  Schr\"odinger representation of $\cA_V$ by the operators
\mar{qq83}\beq
f=a^k(q^\mu)p_k + b(q^\mu), \qquad
\wh f\rho= (-ia^k\dr_k -\frac{i}{2}\dr_k(a^k) +b) \rho
\label{qq83}
\eeq
in the same space of complex half-densities on $Q$ as that of $\cA_T$.
Moreover, this representation preserves the structure of $\cA_V$ as a
$C^\infty(\bR)$-algebra.  

The metaplectic correction of the
Poisson bundle
$V^*Q\to\bR$ also provides the metaplectic corrections of 
its symplectic leaves as follows. It is readily observed that
the Jacobian $J$ restricted to a fibre
$Q_t$, $t\in\bR$, is a Jacobian of coordinate transformations on $Q_t$.
Therefore, any half-density $\rho$ on $Q$, restricted to $Q_t$, is a
half-density on $Q_t$. Then the representation (\ref{qq83}) restricted to a
fibre
$V^*_tQ$ is exactly the metaplectic correction of the Schr\"odinger
representation of the symplectic manifold $V^*_tQ$ on half-densities on $Q_t$.

Thus, the quantum algebra $\cA_V$ of the Poisson bundle $V^*Q\to\bR$,
given by the operators (\ref{qq83}),  can be seen as the
instantwise algebra of operators on its symplectic fibres.  

The representation (\ref{qq83}) can be extended locally to functions on
$T^*Q$ which are polynomials of momenta $p_\la$. These functions can be
represented by elements of the universal enveloping algebra $\ol\cA_T$ of the
Lie algebra $\cA_T$, but this representation is not necessarily globally
defined. For instance, a generic quadratic Hamiltonian 
\mar{qq130}\beq
\cH=a^{jk}(q^\la)p_jp_k + b^k(q^\la)p_k + c(q^\la) \label{qq130}
\eeq
leads to the Hamiltonian function $\cH^*=p+\cH$ which is not an element of
$\ol\cA_T$ because of the quadratic term.  This term can be quantized
only locally, unless the Jacobian of the coordinate transition functions 
on $Q$ is independent of fibre coordinates $q^k$ on $Q$.

\section{Presymplectic quantization}

As was mentioned above, to include a Hamiltonian 
function $\cH^*$ (\ref{mm16}) to the quantum algebra, 
one can choose the
Hamiltonian polarization of $T^*Q$. This polarization accompanies a different
approach to geometric quantization of the momentum phase space of
non-relativistic mechanics $V^*Q$ which is considered as a presymplectic
manifold. In comparison with the above one, this quantization does not lead to
quantization of the Poisson algebra of functions on $V^*Q$ as follows. 

Every global section $h$ (\ref{qq4}) of the affine bundle $\zeta$ (\ref{z11})
yields the pull-back Hamiltonian form
\mar{b4210}\beq
H=h^*\Xi= p_k dq^k -\cH dt  \label{b4210}
\eeq
on $V^*Q$. With respect to a trivialization 
(\ref{gm156}),
the form $H$ is the
well-known  integral invariant of Poincar\'e--Cartan 
\cite{arno}.  Given a Hamiltonian form $H$ (\ref{b4210}), there exists a unique
Hamiltonian connection 
\mar{m57}\beq
\g_H=\dr_t +\dr^k\cH\dr_k -\dr_k\cH\dr^k \label{m57}
\eeq
on the fibre bundle $V^*Q\to \bR$ such that
\mar{qq1}\beq
\g_H\rfloor dH=0 \label{qq1}
\eeq
\cite{book98,sard98}. It defines the Hamilton equations on $V^*Q$.

A glance at the equation (\ref{qq1}) shows that one can think of the
Hamiltonian connection $\g_H$ as being the Hamiltonian vector field of a zero
Hamiltonian with respect to the presymplectic form $dH$ on $V^*Q$. 
Therefore, one can study geometric quantization of the presymplectic manifold
$(V^*Q,dH)$.

Usually, geometric quantization is not applied directly
to a presymplectic manifold $(Z,\om)$, but to a symplectic manifold 
$(Z',\om')$
such that the presymplectic form $\om$ is a pull-back of the symplectic
form $\om'$. Such a symplectic manifold always exists. 
The following two possibilities are usually considered: (i) 
$(Z',\om')$
is a reduction of $(Z,\om)$ along the leaves of the characteristic distribution
of the presymplectic form $\om$ of constant rank \cite{got86,vais83},
and (ii)
there is a coisotropic
imbedding of $(Z,\om)$ to $(Z',\om')$ \cite{got81,got82}.

In application to $(V^*Q,dH)$, the reduction procedure however meets
difficulties. Since the kernel of
$dH$ is generated by the vectors
($\dr_t$, $\dr_k\cH\dr^k-\dr^k\cH\dr_k$, $k=1,\ldots,m$),
the presymplectic form $dH$ in physical models is almost never of
constant rank. Therefore, one has to provide an
exclusive analysis of each physical model, and to 
cut out a certain subset of $V^*Q$
in order to use the reduction procedure.

The second variant of geometric quantization of the presymplectic manifold
$(V^*Q,dH)$ seems more attractive because the section
$h$ (\ref{qq4}) is a coisotropic imbedding.  
Indeed, the tangent bundle $TN_h$ of the closed imbedded submanifold
$N_h=h(V^*Q)$ of $T^*Q$ consists of the vectors
\mar{q2}\beq
u=-(u^\m\dr_\m\cH + u_j\dr^j\cH)\dr^0 + u_k\dr^k + u^\m\dr_\m. \label{q2}
\eeq
Let us prove that the orthogonal distribution Orth$_\Om TN_h$ of $TN_h$ with
respect to the symplectic form $\Om$ belongs to $TN_h$. By definition, it
consists of the vectors
$u\in T_zT^*Q$, $z\in T^*Q$, such that 
\be
u\rfloor v\rfloor\Om=0, \qquad \forall v\in T_zN_h. 
\ee
A simple calculation shows that these vectors obey the conditions
\be
-u^0\dr^i\cH + u^i=0, \qquad u^0\dr_\m-u_\m=0
\ee 
and, consequently, take the form (\ref{q2}).

The image $N_h=h(V^*Q)$ of the coisotropic imbedding $h$ is given by the
constraint
\be
\cH^*=p+\cH(t,q^k,p_k)=0. 
\ee
Then the geometric quantization of the
presymplectic manifold
$(V^*Q,dH)$ consists in geometric quantization of the cotangent bundle
$T^*Q$ and setting the quantum constraint condition
\be
\wh\cH^*\psi =0 
\ee
on physically admissible quantum states. 
This condition implies that $\wh\cH^*$
belongs to the quantum algebra of $T^*Q$. It takes place if the above mentioned
Hamiltonian polarization of
$T^*Q$, which contains the Hamiltonian vector field 
\mar{qq24}\beq
\vt_{\cH^*}=\dr_0 +\dr^k\cH\dr_k -\dr_k\cH\dr^k,  \label{qq24}
\eeq
is used. 

Such a polarization of $T^*Q$ always exists. Indeed, any section
$h$ (\ref{qq4}) of the affine bundle
$T^*Q\to V^*Q$ defines the splitting 
\be
a_\la\dr^\la = a_k(\dr^k -\dr^k\cH\dr^0) +(a_0 +a_k\dr^k\cH)\dr^0 
\ee
of the vertical tangent bundle $VT^*Q$ of $T^*Q\to Q$. One can justify this
fact, e.g., by inspection of the coordinate transformation law. Then elements
$(\dr^k -\dr^k\cH\dr^0$, $k=1,\ldots,m$), and the values of the Hamiltonian
vector field $\vt_{\cH^*}$ (\ref{qq24}) obey the polarization condition
(\ref{qq26}) and generates a polarization of $T^*Q$. It is clear that the
Hamiltonian polarization does not satisfy the condition (\ref{qq14}), and does
not define any polarization of the Poisson manifold $V^*Q$.

Nevertheless, given a trivialization (\ref{gm156}), symplectic fibres
$V_t^*Q$, $t\in \bR$, of the Poisson bundle 
$V^*Q\to\bR$ can be provided with the Hamiltonian polarization $\bT_t$
generated by vectors $\dr_k\cH\dr^k-\dr^k\cH\dr_k$, $k=1,\ldots,m$, except
the points where 
\be
d\cH=\dr_k\cH dq^k + \dr^k\cH dp_k=0.
\ee
This is a standard polarization in conservative Hamiltonian mechanics of
one-dimensional systems, but it requires an exclusive analysis of each model.

\section{Classical and quantum evolution equations}

Turn now to the evolution equation in classical and quantum non-relativistic
mechanics.

Given a 
Hamiltonian connection $\g_H$ on the momentum phase space $V^*Q$,
let us consider the Lie derivative  
\mar{m59}\beq
\bL_{\g_H} f=\g_H\rfloor df=(\dr_t +\dr^k\cH\dr_k -\dr_k\cH\dr^k)f \label{m59}
\eeq
of a function $f\in C^\infty(V^*Q)$ along $\g_H$.
This equality is the evolution equation in classical non-relativistic
Hamiltonian mechanics.  Substituting a solution of the Hamilton equations in
its right-hand side,  one obtains the time evolution of $f$ along this
solution.  Given a trivialization (\ref{gm156}), 
the evolution equation (\ref{m59}) can be written as 
\be
\bL_\g f= \dr_tf +\{\cH,f\}_V. 
\ee
However, taken separately, the terms in its
right-hand side are ill-behaved under time-dependent
transformations. Let us bring the evolution equation
into the frame-covariant form.

The affine bundle $\zeta$ (\ref{z11}) is modelled over 
the trivial line bundle $V^*Q\times \bR\to V^*Q$.
Therefore, Hamiltonian forms $H$ 
constitute an affine space modelled over the vector space 
$C^\infty(V^*Q)$. To choose a centre of this affine space, let us consider
a connection 
\mar{qq73}\beq
\G=\dr_t +\G^k\dr_k \label{qq73}
\eeq
on the configuration bundle $Q\to\bR$. By definition, it 
is a section of the affine bundle (\ref{z11}), and yields the 
Hamiltonian form
\be
H_\G=\G^*\Xi=p_kdq^k -\cH_\G dt, \qquad \cH_\G= p_k\G^kdt.  
\ee
Its Hamiltonian connection is the canonical lift
\be
V^*\G=\dr_t +\G^i\dr_i -p_i\dr_j\G^i \dr^j 
\ee
of the connection $\G$ onto $V^*Q\to \bR$.
Then  any Hamiltonian form $H$ (\ref{b4210}) on the momentum phase space $V^*Q$
admits the splittings
\mar{m46'}\beq
H=H_\G -\wt\cH_\G dt, \qquad \wt\cH_\G=\cH-p_k\G^k, \label{m46'}
\eeq
where $\wt \cH_\G$ is a function on $V^*Q$.
The physical meaning of this splitting becomes clear due to the
fact that
every trivialization of $Q\to \bR$ yields a complete
connection $\G$ on $Q$, and {\it vice versa} \cite{book98,book00}.
From the physical viewpoint, the vertical part of this connection $\G$
(\ref{qq73}) can be seen as a velocity of an "observer", and $\G$  
 characterizes
a reference frame in non-relativistic time-dependent mechanics
\cite{book98,massa,sard98}. 
Then one can show that 
$\wt\cH_\G$ in the splitting (\ref{m46'}) is the energy
function with respect to this reference frame \cite{eche95,book98,sard98}.

Given the splitting (\ref{m46'}), the evolution equation can be written 
in the frame-covariant form
\be
\bL_{\g_H}f= V^*\G\rfloor H +\{\wt\cH_\G,f\}_V. 
\ee
However, the first term in its right-hand side is not reduced to the Poisson
bracket on $V^*Q$, and 
is not quantized in the framework of geometric quantization of the
Poisson manifold $V^*Q$.
 To bring the right-hand side of the evolution equation into a Poisson
bracket alone, let us consider the pull-back
$\zeta^*H$ 
of the
Hamiltonian form
$H=h^*\Xi$ onto the cotangent bundle $T^*Q$. It is readily observed that the
difference
$\Xi-\zeta^*H$ is a horizontal 1-form on $T^*Q\to\bR$, and we obtain the
function (\ref{mm16}) on $T^*Q$. Then the relation
\be
\zeta^*(\bL_{\g_H}f)=\vt_{H^*}(\zeta^*f) =\{\cH^*,\zeta^*f\}_T, 
\ee
holds for any element $f$ of the Poisson algebra $C^\infty(V^*Q)$
\cite{jmp00}. 

Since the quantum algebra $\cA_V$ of the Poisson manifold $V^*Q$ can
be seen as the instantwise algebra, one can quantize the evolution equation
(\ref{mm17}) as follows. 

Given the Hamiltonian function $\cH^*$ (\ref{mm16}), let $\wh \cH^*$ be the
corresponding quantum operator written, e.g., as an element of universal
enveloping algebra of the quantum algebra $\cA_T$. 
The bracket 
(\ref{qq120})
defines a derivation of the enveloping algebra $\ol\cA_V$ of the quantum
algebra
$\cA_V$, which is also a $C^\infty(\bR)$-algebra. Moreover, since $\wh
p=-i\dr/\dr t$, the derivation (\ref{qq120}) obeys the Leibniz rule
\be
\nabla (r(t)\wh f)=\dr_t r(t)\wh f + r(t)\nabla \wh f.
\ee  
Therefore, it is a connection on the $C^\infty(\bR)$-algebra $\ol\cA_V$, and
defines quantum evolution of $\ol\cA_V$ as a parallel transport
along time \cite{book00,sard00}.


\begin{thebibliography}{ederf}

\bibitem{arno} V.Arnold, {\it Mathematical Methods of Classical Mechanics}
(Springer, Berlin, 1978).

\bibitem{eche95} A.Echeverr\'{\i}a Enr\'{\i}quez, M.Mu\~noz Lecanda and
N.Rom\'an Roy, Non-standard connections in classical mechanics, 
{\it J. Phys. A} {\bf 28} (1995) 5553.

\bibitem{eche98} A.Echeverr\'{\i}a Enr\'{\i}quez, M.Mu\~noz Lecanda, 
N.Rom\'an Roy, and C.Victoria-Monge, Mathematical foundations of geometric
quantization, {\it Extracta Math.} {\bf 13} (1998) 135.

\bibitem{got81} M.Gotay and J.\'Sniatycki, On the quantization of
presymplectic dynamical systems via coisotropic imbeddings, {\it Commun.
Math. Phys.} {\bf 82} (1981) 377.

\bibitem{got82} M.Gotay, On coisotropic imbeddings of presymplectic
manifolds, {\it Proc. Amer. Math. Soc.} {\bf 84} (1982) 111.

\bibitem{got86} M.Gotay, Constraints, reduction and quantization, {\it J.
Math. Phys.} (1986) 2051. 

\bibitem{leon} M. de Le\'on, J.Marrero and E.Padron, On the geometric
quantization of Jacobi manifolds, {\it J. Math. Phys.} {\bf 38} (1997) 6185.

\bibitem{book98} L.Mangiarotti and G.Sardanashvily, {\it Gauge Mechanics}
(World Scientific, Singapore, 1998).

\bibitem{book00} L.Mangiarotti and G.Sardanashvily, {\it Connections in
Classical and Quantum Field Theory} (World Scientific, Singapore, 2000).

\bibitem{jmp00} L.Mangiarotti and G.Sardanashvily, Constraints in Hamiltonian
time-dependent mechanics, {\it J. Math. Phys.} {\bf 41} (2000) 2858

\bibitem{massa} E.Massa and E.Pagani, Jet bundle geometry, dynamical
connections and the inverse problem of Lagrangian mechanics, {\it Ann. Inst.
Henri Poincar\'e} {\bf 61} (1994) 17.

\bibitem{sard98} G.Sardanashvily, Hamiltonian time-dependent mechanics,
{\it J. Math. Phys.} {\bf 39} (1998) 2714.

\bibitem{sard00} G.Sardanashvily, Classical and quantum mechanics with
time-dependent parameters, {\it J. Math. Phys.} {\bf 41} (2000) 5245.

\bibitem{sni} J.\'Sniatycki, {\it Geometric Quantization and Quantum
Mechanics} (Springer-Verlag, Berlin, 1980).

\bibitem{vais83} I.Vaisman, Geometric quantization on presymplectic manifold,
{\it Monatsh. Math.} {\bf 96} (1983) 293.

\bibitem{vais91} I.Vaisman, On the geometric quantization of Poisson
manifolds, {\it J. Math. Phys.} {\bf 32} (1991) 3339.

\bibitem{vais} I.Vaisman, {\it Lectures on the Geometry of Poisson Manifolds}
(Birkh\"auser Verlag, Basel, 1994).

\bibitem{vais97} I.Vaisman, On the geometric quantization of the symplectic
leaves of Poisson manifolds, {\it Diff. Geom. Appl.} {\bf 7} (1997) 265.

\bibitem{wood} N.Woodhouse, {\it Geometric Quantization} (Clarendon Press,
Oxford, 1980) (2$^{\rm nd}$ ed. 1992).


\end{thebibliography}
\end{document}